\documentclass[12pt]{iopart}

\usepackage{iopams}  
\usepackage{setstack}

\newcommand{\ket}[1]{\left\vert#1\right\rangle}
\newcommand{\bra}[1]{\left\langle#1\right\vert}

\newcommand{\dmat}[2]{\ket{#1}\!\!\bra{#2}}
\newcommand{\beq}{\begin{equation}}
\newcommand{\eeq}{\end{equation}}
\newcommand{\bea}{\begin{eqnarray}}
\newcommand{\eea}{\end{eqnarray}}

\makeatletter
\def\btt#1{\texttt{\@backslashchar#1}}
\DeclareRobustCommand\bblash{\btt{\@backslashchar}}
\makeatother
\def\Bid{{\mathchoice {\rm {1\mskip-4.5mu l}} {\rm
{1\mskip-4.5mu l}} {\rm {1\mskip-3.8mu l}} {\rm {1\mskip-4.3mu l}}}}

\begin{document}

\title{Compatible Transformations for a Qudit 
Decoherence-free/Noiseless Encoding}

\author{C. Allen Bishop}

\address{Physics Department, Southern Illinois University, 
Carbondale, Illinois 62901-4401}

\ead{abishop@www.physics.siu.edu}

\author{Mark S. Byrd}

\address{Physics Department and Computer Science Department, 
Southern Illinois University, Carbondale, Illinois 62901-4401}

\ead{mbyrd@physics.siu.edu}

\begin{abstract}
The interest in decoherence-free, or noiseless subsystems (DFS/NSs)
of quantum systems is both of fundamental and practical interest. 
Understanding the invariance of a set of states under certain transformations 
is mutually associated with a better understanding of 
some fundamental aspects of quantum
mechanics as well as the practical utility of invariant 
subsystems.  For example, DFS/NSs are potentially useful for protecting quantum
information in quantum cryptography and quantum computing as well as
enabling universal computation.  
Here we discuss transformations which are compatible 
with a DFS/NS that is composed of $d$-state systems which protect
against collective noise. They are compatible in 
the sense that they do not take the logical (encoded) 
states outside of the DFS/NS during the transformation. Furthermore, it is 
shown that the Hamiltonian evolutions derived here can be used 
to perform 
universal quantum computation on a three qudit DFS/NS. Many of the
methods used in our derivations are directly
applicable to a large variety of DFS/NSs.  More generally, we may also state 
that these transformations are compatible with collective 
motions.  
\end{abstract}




\maketitle


\section{Introduction}

In 1954, Dicke argued that an independent approximation for 
radiating physical systems was very often not reasonable 
\cite{Dicke:54}.  Many physical systems are correlated, 
and transform in a similar, or even identical way  
so that an independent approximation should not be used. 
For some systems the motions can be considered collective.  
For example, this can be the case when a set of particles 
interacts with a common field.  In 
proposed quantum computing devices, interactions 
with a common field can lead to collective noise which is an 
{\sl unwanted} interaction.  
This collective noise, or collective decoherence, 
will cause quantum information to be 
lost to the environment \cite{Zurek:03,Nielsen/Chuang:book}.  

To avoid the detrimental effects of such noise, a theory of 
decoherence-free or noiseless subsystems (DFS/NSs) was formulated 
\cite{Zanardi:97c,Duan:98,Lidar:PRL98,Knill:99a,Kempe:00,Lidar:00a}.   
(For reviews see \cite{Lidar/Whaley:03,Byrd/etal:pqe04}.) 
When information is encoded into one particular type of DFS/NS it 
is protected against collective errors.   
More generally, under suitable conditions a 
DFS/NS can also protect against noises which are not of the collective
type, but still correspond to an identifiable symmetry in the
system-bath interaction.  
However, in this article we will focus on collective DFS/NSs. Once 
information is encoded into a DFS/NS, it must then 
be manipulated if the quantum information is to be used 
for the purposes of computation or simulation.  Not just any 
physically available operation is acceptable.  
The manipulations, or gating operations, 
must be {\it compatible} with the DFS/NS \cite{Kempe:00} if 
the information is to
remain protected.  
Here compatible means that the operations should not take 
the information outside of the DFS/NS.  For, if they did, 
the information would be vulnerable to collective noises 
during the time it is not confined to the DFS/NS.  

The objective of this paper is to describe Hamiltonian 
evolutions which are compatible with collective operations 
on the system in the sense that they commute with the 
collective operations.  The Hamiltonians we find here will 
provide a set of transformations which are compatible with 
a DFS/NS that protects information from collective noise 
and is composed of qudits.  

Qudits are quite interesting systems for several reasons.  
For example, two three-state systems, or qutrits, can be more 
entangled than two qubits
\cite{Caves/Milburn:99,Rungta/etal:00,qutritent}. There 
is also evidence suggesting that a collection of $d$-state systems 
can share a larger pairwise fraction of 
their entanglement capacity as the dimension 
increases \cite{Wootters:dqudits}.  They are
beneficial for several information processing 
tasks including cryptography 
\cite{Pasquinucci/etal:June2000,Pasquinucci/etal:Oct2000,%
Bourennane/etal:01,Bruss/etal:02,Durt/etal:03}, computing 
\cite{Bartlett/etal:02,Klimov/etal:03,Ralph/etal:07}, 
and games.  They seem to be {\sl required} for a version 
of the Byzantine agreement problem \cite{Fitzi/etal:01}.  
So for these reasons, among others, universality requirements for qudits 
\cite{Muthukrishnan:00,Kempe:01e,BBO:05,Brennen/etal:05,Bullock/etal:05,Bacon/etal:07}
as well as several recent experiments 
and proposed experiments have provided methods for 
producing and manipulating qudits \cite{Gaertner/etal:08,Bishop/Byrd:08,Vallone/etal:07,
Bogdanov/etal:04} and an analysis of the collective behavior for an 
assembly of qutrits has been given \cite{Mandilara/Akulin:07}.

Collective DFS/NSs have been 
studied quite thoroughly for a variety of reasons.  One is 
that collective errors are physically observed in some systems.  
Another is that a collective DFS encoding can enable 
universal computation on a set of qubits 
 \cite%
{Bacon:99a,Kempe:00,Bacon:Sydney,DiVincenzo:00a,Levy:01,%
Benjamin:01,Wu/Lidar:01a,Lidar/Wu:01,Kempe:01d,Kempe:01e,Lidar/Wu/Blais:02}. Collective 
DFS/NSs have also been observed to reduce noise in several experiments 
\cite{Kwiat:00,Kielpinski:01,Viola/etal:01,Fortunato/etal:02,Altepeter:04}, 
including computation in a 
qubit DFS \cite{Ollerenshaw:03,Mohseni:03}.  
Even if collective errors are not present in an experiment, it
is possible to induce such a symmetry using decoupling operations 
\cite{Zanardi:98b,Viola:00a,Wu/Lidar:cdfs,Byrd/Lidar:ss,Byrd/Lidar:ebb,Viola:01a}.  
Theoretically collective noises are easier to treat in part because 
a basis for 
collective operations forms a representation of the algebra of the 
special unitary group \cite{Byrd:06} acting on the constituents.  
This enables a variety 
of group-theoretical methods to be employed in their treatment.  
We will see here that this is also the case for computing in 
a qudit DFS/NS.   

Specifically in this paper, Section \ref{sec:dfss} contains 
a review of DFS/NS theory.  We then provide a description 
of the compatibility condition and show how it can be used 
to find a set of Hamiltonians which enable compatible 
computation on a DFS/NS in Section \ref{sec:comp}.  
Section \ref{sec:Hams} contains a set of compatible Hamiltonians for
a system of three qudits.  We provide a complete set for qutrits 
in Section \ref{sec:trits} and then generalize to qudits.  
We then show how to analytically obtain several unitary
transformations on encoded states using these 
results in Section \ref{sec:unicomp}.  Section \ref{sec:ndits} extends 
this analysis to a system of $n$-qudits ($n \geq 3$), thereby showing 
that these compatible Hamiltonians can be used to manipulate 
encoded quDit states ($D \neq d$) of arbitrary dimension. We conclude in 
Section \ref{sec:concl} with a discussion of our results 
and their implication for universality using a three qudit DFS/NS. 
Two appendices have also been included which contain some detailed 
calculations.


\section{Decoherence-Free/Noiseless Subsystems}

\label{sec:dfss}

Here a brief review of DFS/NSs is provided.  Although the description is, 
by now, more or less standard, we will primarily follow the notation of 
Ref.~\cite{Kempe:00}.  For further details, see \cite{Knill:99a} and/or
\cite{Kempe:00}.  We note that one may describe 
a DFS/NS in terms of a collection of operators 
appearing in a semi-group master equation \cite{Lidar:PRL98}.  
We will, however, use the Hamiltonian description here.


\subsection{Subsystem Structure}

Consider a Hamiltonian which includes a term describing the evolution
of the system ($H_S$), a term describing the evolution of a bath or
environment ($H_B$), and an interaction term describing the evolution 
of the system and environment together, ($H_{SB}$)
\beq
H = H_S + H_B +H_{SB}.
\eeq
Without loss of generality, we will write the interaction 
term as 
\beq
H_{SB}=\sum_\alpha S_\alpha\otimes B_\alpha,
\eeq 
where the $S_\alpha$ act only on the system and the $B_\alpha$ act only on
the bath.  Let us now consider the algebra ${\cal A}$ generated by the set of
operators $\{H_S,S_\alpha \}$.  We will assume that this algebra is reducible
and thus the operators can all be simultaneously block-diagonalized
with a single unitary transformation.  Such algebras are said to be,
mathematically, completely reducible, meaning they may be completely
block diagonalized, i.e., written as a direct sum of irreducible
components.  It is this reducibility, which 
follows for any system-bath interaction symmetry that is preseved in
the algebra, that allows for a non-trivial decomposition of the
algebra into regions supporting the logical encoding of protected
information.  It is noteworthy that the ``unitary trick'' can be used to
relate algebraic representation theory and group representation
theory \cite{Byrd:06}.  This is particularly useful for collective errors acting on 
systems of qudits since these operations 
form a representation of the algebra of such systems. 

This decomposition is described by the
equation
\begin{equation}
\label{eq:Adef}
{\cal A} \cong \underset{J\in {\cal J}}{\bigoplus}\Bid_{n_J}\otimes 
                                     {\cal M}(d_J,\mathbb{C}),
\end{equation}
where the $n_J$-fold degenerate $d_J \times d_J$ complex matrices 
${\cal M}(d_J,\mathbb{C})$ 
correspond to the irreducible components of 
${\cal A}$. We label these components by $J$, which collectively form the finite 
set ${\cal J}$.  (It is important to note that this $J$ actually
stands for a set of quantum numbers when the constituents are qudits, with
$d\geq 3$.)  
One may also define the commutant which is the set of elements that
commute with all elements of the algebra ${\cal A}$.  The form of these is
dictated by the irreducibility of the blocks in ${\cal A}$,
\begin{equation}
{\cal A}^\prime = \{X:[X,A]=0, \forall A\in {\cal A}\}.  
\end{equation}
To Eq.~(\ref{eq:Adef}) there is a corresponding decomposition of
the Hilbert space ${\cal H}_{S} =\sum_J \mathbb{C}^{n_J} \otimes
\mathbb{C}^{d_J}$ where the second factor corresponds to the part of the
Hilbert space which is affected by noise ($\mathbb{C}^{d_J}$) and the 
first factor corresponds to that part which is not ($\mathbb{C}^{n_J}$).  

The unitary transformation which is used to 
to change between the physical and logical bases 
can be referred to as the DFS/NS transformation. In the logical basis, 
the elements of ${\cal A}$ 
exhibit a structure that allows quantum information to remain confined 
to the logical subspaces while the physical system interacts with 
its environment. The superpositions of physical states which form the logical, or encoded,
states of the DFS/NS are created by this transformation in the following way.
Let $V_{\mbox{\scriptsize dfs}}$ be the 
aforementioned transformation that simultaneously block diagonalizes each element of the algebra
$A_i\in{\cal A}$ in an identical way, i.e., $A_i^{\prime} =
V_{\mbox{\scriptsize dfs}}^{\phantom{A}}A_i V_{\mbox{\scriptsize dfs}}^{-1}$. The physical states $\ket{\Psi_p}$ are then
related to the logical states $\ket{\Psi_L}$ by
\beq
\ket{\Psi_L} = V_{\mbox{\scriptsize dfs}}\ket{\Psi_p}.
\eeq
In practice, $V_{\mbox{\scriptsize dfs}}$ can be used to express states and 
operators in the physical bases in terms of states and operators in
the logical basis as done explicitly, for example, in \cite{Byrd/etal:05}.

We may now define a decoherence-free or noiseless {\it subsystem} in the
following way.  Suppose that we represent a basis of eigenstates 
corresponding to a particular $J$ with the set 
$\{\ket{\lambda} \otimes \ket{\mu}\}$, where 
$\lambda = 1,...,n_J$ and $\mu = 1,...,d_J$. Then if 
\beq
\label{eq:dfsalg2}
A_\alpha\ket{\lambda}\otimes\ket{\mu} = \sum_{\mu^\prime =1}^{d_J}
    M_{\mu\mu^\prime,\alpha}\ket{\lambda}\otimes\ket{\mu^\prime}
\eeq
for all $A_\alpha, \lambda,$ and $\mu$, there exists an 
irreducible decomposition as given by Eq.~(\ref{eq:Adef}). The invariance 
of the degeneracy labels $\ket{\lambda}$ in this last expression 
reflects the ability to reliably store quantum information in 
certain regions of the system Hilbert space when the algebra ${\cal A}$ 
can be decomposed in this way. The information is stored in blocks with
the same $J$ but different $\lambda$.  Each $\lambda$ specifies 
a particular DFS/NS basis state. These logical states can therefore be 
expanded in terms of those states associated with a given $\lambda$. 
Although the initial encoding of a particular logical state may
change, it will remain confined to its initial subspace.  
 
A decoherence-free {\it subspace} is one for whcih the
matrices $M$ are numbers (1 $\times$ 1 matrices) which act on a
one-dimensional representation, i.e., a singlet state.


\subsection{DFS and NS examples}

In this section we review a few examples of DFS/NSs which 
will aid in our discussion of qudit systems.  
The examples will be useful for comparing and contrasting 
certain properties of qubit and qudit systems. 


\subsubsection{Four-qubit DFS}

As stated above a decoherence-free subspace is comprised of 
singlet states.  Four qubits can be used to 
construct a DFS/NS qubit which is represented by two singlet 
states, one singlet state for the logical zero and one for the logical
one \cite{Lidar:PRL98}.  The logical states considered here protect
the information from errors which act the same on each of the four
physical qubits constituting the system. These states are given
explicitly by
\begin{equation}
\ket{0_L}=(\ket{0101}+\ket{1010}-\ket{0110}-\ket{1001})/2,
\end{equation}
and
\bea
\ket{1_L}&=& (2\ket{0011}+2\ket{1100}-\ket{0110} \nonumber \\
&&-\ket{1001}-\ket{0101}-\ket{1010})/\sqrt{12},
\eea
where $\ket{0}$ and $\ket{1}$ represent two orthogonal basis 
states for a spin-1/2 particle. 

It turns out that the Heisenberg exchange interaction is universal 
for a set of qubits constructed in this way and 
logical gates consisting of only these interactions have been provided 
\cite{Bacon:99a,Kempe:00,Bacon:Sydney,Hsieh:03}. 
(In this paper we neglect the corrections which must be 
made to the logical gates in order to account for three- and four-body 
interaction terms. For a discussion of these effects see, for example, 
Ref.~\cite{Woodworth:06}.) The Heisenberg exchange interaction between pairs of physical 
qubits can be expressed as:%
\begin{equation}
E_{ij}=\frac{1}{2}(I+\vec{\sigma }_{i}\cdot \vec{\sigma }_{j}),
\end{equation}%
where $\vec{\sigma }=(\sigma _{x},\sigma _{y},\sigma _{z})$ is the vector of
Pauli matrices. As written $E_{ij}$ is the exchange operation between qubits 
$i$ and $j$, i.e., $E_{ij}|\phi \rangle _{i}|\psi \rangle _{j}=|\psi \rangle
_{i}|\phi \rangle _{j}$.               
The logical ``$X$'' operation is given by 
\begin{equation}
\label{eq:4qbXgate}
\bar{X}=\frac{1}{\sqrt{3}}(E_{23}-E_{13}).
\end{equation}
The logical ``$Z$'' operation is given by 
\begin{equation}
\bar{Z}=-E_{12}
\end{equation}
and $\bar{Y}$ can be obtained from these two by commutation or the 
finite transformations can be written in terms of Euler angles. 


\subsubsection{Three-qubit DFS}

The smallest number of qubits that enables the DFS/NS 
encoding of a logical qubit is three.  In this case the 
logical zero and logical one are each represented by a 
doublet. In the $\{\ket{0},\ket{1}\}$ basis, these states can 
be written as
\beq
\ket{0_L}=\alpha_0(\ket{010}-\ket{100})/\sqrt{2}
+\beta_0(\ket{011}-\ket{101})/\sqrt{2},
\eeq
and
\bea
\ket{1_L}&=&\alpha_1(2\ket{001}-\ket{010}-\ket{100})/\sqrt{6} \nonumber \\
&&+\beta_1(-2\ket{110}+\ket{011}+\ket{101})/\sqrt{6},
\eea
with $|\alpha_i|^2 + |\beta_i|^2 = 1$. 
As in the previous example, these logical states protect 
the quantum information from collective errors.  
Although the Heisenberg exchange interaction is universal for 
both the three-qubit and four-qubit DFS/NSs, there are 
some practical differences between the respective logical operations for these systems 
with regard to error prevention \cite{Wu/etal:02,Byrd/etal:05}.  
We will not discuss this here, but will note the form of 
the logical operations.  The logical 
``$X$'' operation is given by \cite{Kempe:00} 
\begin{equation}
\label{eq:3qbXgate}
\bar{X}=\frac{1}{\sqrt{3}}(E_{23}-E_{13}),
\end{equation}
while the logical ``$Z$'' operation is given by \cite{Kempe:00-fix} 
\begin{equation}
\label{eq:3qbZgate}
\bar{Z}=\frac{1}{3}(E_{13}+E_{23}-2E_{12}).
\end{equation}%
Again, $\bar{Y}$ can be obtained from these two by commutation or 
Euler angles.

It is important to note the similarities and differences 
here.  However, for our purposes, the most important point 
is that the Heisenberg exchange interaction can be used to 
construct a universal set of operations for both the 
three- and four-qubit DFS.


\subsubsection{Three-qutrit NS}

Here we provide some details concerning the logical qubit 
encoding over a subspace of three physical qutrits. The purpose of 
this particular example is to provide enough structure to 
enable a smooth transition into our discussion of qudit systems.

The three qutrit DFS/NS is quite analogous to the three qubit 
DFS/NS discussed above in many respects although the dimensions 
of the subspaces are different.  For the three qubit 
NS, a tensor product of three qubits can be decomposed into 
two two-state subsystems and a four-state system.  This may 
be written as ${\bf 2}\otimes{\bf 2}\otimes{\bf 2} = %
{\bf 2}\oplus{\bf 2}\oplus{\bf 4}$.  
A tensor product of three qutrits can be decomposed 
into a singlet, a decuplet, and two octets: 
\begin{equation}
\label{eq:decomposition}
{\bf 3}\otimes {\bf 3}\otimes {\bf 3} = 
{\bf 8}\oplus {\bf 8} \oplus {\bf 1} \oplus {\bf 10}, 
\end{equation}
where the 
${\bf 8}$ identifies an eight-dimensional representation (octet), the  
${\bf 1}$ identifies a one-dimensional representation (singlet), etc. 
A logical qubit can be represented by two 
degenerate eight-state subsystems, the two octets \cite{Byrd:06}. 
We may also note that it is possible to experimentally 
produce such states encoded in a set of polarized 
photons \cite{Bishop/Byrd:08}.

Explicit forms for the states within 
the two octet subspaces of three physical qutrits were provided in 
\cite{Byrd:06}.  
The three states within each individual qutrit will be denoted 
$\ket{0}$, $\ket{1}$, or $\ket{2}$.  
In terms of the quantum numbers $p,q,t,t_3,$ and $y$ as given in 
\cite{Byrd:06}, these are \cite{note:basis}
\begin{eqnarray}
\label{eq:3states}
\ket{0} &=& \ket{0,1,-1/2,-1/2,-1/3},\nonumber \\
\ket{1} &=& \ket{0,1,-1/2,1/2,-1/3},\\
\ket{2} &=& \ket{0,1,0,0,2/3}. \nonumber
\end{eqnarray} 
The physical, or computational basis ($\ket{000}$, $\ket{001}$, ... , 
$\ket{222}$) transforms into the DFS basis via the operation 
$V_{\mbox{\scriptsize dfs}}$ which is the matrix composed of 
the Wigner-Clebsh-Gordan coefficients.  
In this noiseless basis the logical zero state $\ket{0_L}$ is formed 
from the octet which corresponds to the states \cite{Byrd:06}
\begin{eqnarray}
\label{eq:octet1}
\psi_1^{8,0} &=& (\ket{200} - \ket{020})/\sqrt{2}, \nonumber \\
\psi_2^{8,0} &=& (\ket{100} - \ket{010})/\sqrt{2}, \nonumber \\
\psi_3^{8,0} &=& (\ket{011} - \ket{101})/\sqrt{2}, \nonumber \\
\psi_4^{8,0} &=& (\ket{211} - \ket{121})/\sqrt{2},  \\
\psi_5^{8,0} &=& (\ket{122} - \ket{212})/\sqrt{2}, \nonumber \\
\psi_6^{8,0} &=& (\ket{022} - \ket{202})/\sqrt{2}, \nonumber \\
\psi_7^{8,0} &=& (-\ket{021} - \ket{120} +  \ket{201} +  \ket{210})/2, 
\nonumber \\
\psi_8^{8,0} &=& (2\ket{012} + \ket{021} - 2\ket{102} \nonumber \\
&& -\ket{120} - \ket{201} + \ket{210})/\sqrt{12}, \nonumber
\end{eqnarray}
where the first superscript on each $\psi$ denotes the dimension 
of the representation, the second is a degeneracy label and the 
subscript labels the state within the representation. 
A second octet forms the logical one state $\ket{1_L}$,
\begin{eqnarray}
\label{eq:octet2}
\psi_1^{8,1} &=& (-2\ket{002} + \ket{020} + \ket{200})/\sqrt{6}, \nonumber \\
\psi_2^{8,1} &=& (-2\ket{001} + \ket{010} + \ket{100})/\sqrt{6}, \nonumber \\
\psi_3^{8,1} &=& (-2\ket{110} + \ket{011} + \ket{101})/\sqrt{6}, \nonumber \\
\psi_4^{8,1} &=& (-2\ket{112} + \ket{121} + \ket{211})/\sqrt{6},  \\
\psi_5^{8,1} &=& (-2\ket{221} + \ket{122} + \ket{212})/\sqrt{6}, \nonumber \\
\psi_6^{8,1} &=& (-2\ket{220} + \ket{022} + \ket{202})/\sqrt{6}, \nonumber \\ 
\psi_7^{8,1} &=& (-2\ket{012} + \ket{021} - 2\ket{102}  \nonumber \\
&& +\ket{120} +\ket{201} +\ket{210})/\sqrt{12}, \nonumber \\
\psi_8^{8,1} &=& (\ket{021} - \ket{120} +\ket{201} -\ket{210})/2. \nonumber
\end{eqnarray}

In terms of these octets the logical zero state is given by an 
arbitrary superposition of the eight $\psi_j^{8,0}$ states, 
$\ket{0_L} = \sum_{j}\alpha_{j}\psi_j^{8,0}$ and likewise for 
$\ket{1_L} = \sum_{j}\beta_{j}\psi_j^{8,1}$. 
According to the theory of noiseless subsystems, states within 
the sets $\{\psi_i^{8,0}\}$  ($\{\psi_i^{8,1}\}$) will mix together 
in the presence of collective noise but not with the states 
in $\{\psi_i^{8,1}\}$ ($\{\psi_i^{8,0}\}$) or with 
the one- or ten-dimensional representations. Furthermore, the mixing 
will be identical for both octets. This means that when only
collective errors are present, the information encoded 
in $\ket{\psi_L} =a \ket{0_L} + b\ket{1_L}$ will be protected.  
It is also important to note that 
the initialization is arbitrary in the sense that an arbitrary 
combination can be taken \cite{Shabani/Lidar:05}.  In 
practical situations this can be very beneficial.  (See 
for example \cite{Bishop/Byrd:08}.)


\subsubsection{Three-qudit DFS}

As discussed in \cite{Byrd:06} 
the tensor product of three equivalent irreducible 
representations of SU($d$) gives rise to the smallest number of 
qudits for which a NS, representing a qubit in terms of qudits,  
exists.  This can be seen in 
the tensor product of three $d$-state systems   
using the tableau of a representation.  
Taking the tensor product of three such systems 
produces two degenerate tableau.  These correspond to 
two degenerate irreducible representations which can 
be used to store protected quantum information in the form 
of a noiseless qubit.  

As we will see later, this decomposition is valid for 
all three-qudit systems and 
enables us to calculate the compatible transformations 
governing the noiseless evolution of such systems.


\section{Computing in a DFS/NS}
\label{sec:comp}

When information is encoded in a DFS/NS, it will 
be protected during the unitary transformations which are used to 
manipulate it only if those transformations 
do not couple states inside the subspace with states outside 
of the subspace.  In other words, we must restrict the transformations 
to those which preserve the subspace structure.  
When this can be accomplished, we say that such logical operations 
are {\it compatible} with the DFS/NS.  In order to find a 
DFS-compatible set of gates we will find it useful to 
first identify the stabilizer of a DFS.


\subsection{The Stabilizer of a DFS/NS}

Let $\ket{\Psi} \in {\cal C}$ be a state in the DFS, 
where ${\cal C}$ denotes the code space.  
Let us define a stabilizer, ${\cal S}$, as in 
\cite{Kempe:00}:
$$
{\cal S} = \{S:S\ket{\Psi}=\ket{\Psi}, \forall \ket{\Psi}\in {\cal C}\}.
$$
While the elements of this set do in fact leave all of the code words 
unchanged, we can relax this requirement slightly when considering 
the DFS states since in this case the information being encoded 
is stored in the states labeled by $\lambda$. This property of 
the DFS/NS states allows us to define a modified version of the 
stabilizer above as
$$
{\cal S}^{\prime} = \{S^{\prime}:S^{\prime}(\ket{\lambda} 
\otimes \ket{\mu})=\ket{\lambda} \otimes \sum_{\mu^{\prime}=1}^{d_J}
C_{\mu \mu^{\prime}}\ket{\mu^{\prime}}, \forall 
\ket{\Psi}\in {\cal C}\}.
$$
We can parametrize 
elements $S^{\prime}$ of the modified stabilizer which are relevant 
to the system using 
Eq.~(\ref{eq:dfsalg2}) and a set of arbitrary complex numbers $\{v_\alpha\}$:
\beq
\label{eq:stab1}
D(v_1,v_2, ...) = \exp\left[\sum_\alpha v_\alpha A_\alpha \right]
\eeq
This leaves the labels $\lambda$ unchanged, as seen by 
expanding the exponential and acting term by term on states 
of the form $\ket{\lambda}\otimes\ket{\mu}$. 


\subsection{Compatibility Conditions for DFS/NS Evolution}

Using the modified stabilizer, 
we can state that $U$ is compatible 
with ${\cal C}$, if $\forall \ket{\Psi}\in {\cal C}$, 
$U(\ket{\lambda} \otimes \ket{\mu}) = \sum_{\lambda^{\prime}}
J_{\lambda \lambda^{\prime}}\ket{\lambda^{\prime}} 
\otimes \sum_{\mu^{\prime}}K_{\mu \mu^{\prime}}\ket{\mu^{\prime}}$ 
and therefore 
$S^{\prime}U(\ket{\lambda} \otimes \ket{\mu}) = S^{\prime}
(\sum_{\lambda^{\prime}}
J_{\lambda \lambda^{\prime}}\ket{\lambda^{\prime}} 
\otimes \sum_{\mu^{\prime}}K_{\mu \mu^{\prime}}\ket{\mu^{\prime}}) = 
(\sum_{\lambda^{\prime}}
J_{\lambda \lambda^{\prime}}\ket{\lambda^{\prime}} 
\otimes \sum_{\mu^{\prime}}K^{\prime}_{\mu \mu^{\prime}}\ket{\mu^{\prime}})$ 
$\forall S^{\prime}\in {\cal S^{\prime}}$. This implies that 
$U^{-1}S^{\prime}U(\ket{\lambda} \otimes \ket{\mu}) = 
\ket{\lambda} \otimes 
\sum_{\mu^{\prime}}K^{\prime \prime}_{\mu  \mu^{\prime}}\ket{\mu^{\prime}}$ so that
\beq
\label{eq:stabcond}
U^{-1}S^{\prime}U \in {\cal S^{\prime}}.  
\eeq
The condition Eq.~(\ref{eq:stabcond}) may now be re-expressed as 
\beq
\label{eq:uscond}
UD(v_1,v_2, ...)U^\dagger = S^{\prime},
\eeq
for some $S^{\prime} \in {\cal S^{\prime}}$, or  
\bea
U  \exp\left[\sum_\alpha v_\alpha A_\alpha\right]U^\dagger 
     =  \exp\left[\sum_\alpha v_\alpha UA_\alpha U^\dagger\right] 
     =   S^{\prime}.
\eea
Taking the natural logarithm of this equation gives 
\beq
\sum_\alpha v_\alpha UA_\alpha U^\dagger 
     =   \ln(S^{\prime}).  
\eeq
Now let us define $U = \exp\left[-iHt\right]$. Then, after taking 
the derivative 
of both sides of this last expression with respect to time, 
we find a very simple and sufficient, but not necessary condition for 
the compatibility of Hamiltonians
\beq
\label{eq:HScond}
[H,A_{\alpha}]=0, \; \forall A_{\alpha}\in{\cal S^{\prime}}.
\eeq
Then $U= \exp(-iHt)$ is compatible with the DFS/NS.  
Clearly if the Hamiltonian which generates the unitary $U$ commutes 
with every element of the algebra, then $U$ also commutes with 
every element of the algebra.  This provides a subset of the 
unitary transformations $U$ satisfying Eq.~(\ref{eq:stabcond}).


\section{Explicit forms for Compatible Hamiltonians} 
\label{sec:Hams}

It was shown in the previous section how a parametrization 
of the stabilizer in terms of time independent coefficients 
$v_\alpha$ can be used to identify a set of unitary 
transformations that are compatible with the 
DFS/NS states. This section provides a set of those Hamiltonians 
satisfying Eq.~(\ref{eq:HScond}) for systems composed 
of three qudits.  In the following analysis we neglect the internal 
Hamiltonian of the system and assume 
that the only interaction terms present are those which 
can be considered as collective. The set presented here is valid 
for all $d \geq 3$ and it is unique for the case of $d = 3$.  
The algorithm for determining this set could be used for any DFS/NS,
not only those which protect against collective errors.  


\subsection{Explicit forms for Qutrits} 

\label{sec:trits}

Using {\texttt{MATHEMATICA}}, we have determined the complete 
set of Hamiltonians which are compatible with an encoded 
two-state noiseless subsystem of three qutrits (compatible in 
the sense that they satisfy Eq.~(\ref{eq:HScond})).  
To do this, we expanded the relevant quantities in a complete 
set of traceless, Hermitian matrices $\{\lambda_i\}$.  For these 
calculations, we employed the Gell-Mann matrices, but 
in principle any basis can be used.  Several properties and 
conventions of these matrices are given in \ref{app:sunalg}.

Collective errors will be denoted $S_\alpha$, of which 
there are eight.  For example, $S_1 = \sum_i\lambda^{(i)}_1$, 
where the superscript labels a particular qutrit and the subscript 
denotes the type of error labeled 1 through 8.  
Let $\mu_{ijk} = \lambda_i\otimes\lambda_j\otimes\lambda_k$, 
where the $\lambda_j$ are the Gell-Mann matrices with 
$\lambda_0=\Bid$.  
An arbitrary Hamiltonian can be expanded as 
$H = \sum_{ijk}a_{ijk}\mu_{ijk}$ with $H$ traceless if 
not all $i,j,k$ are simultaneously zero and the $a_{ijk}$ 
are arbitrary real constants.  The algorithm we used to find 
the $H$ which commutes with every collective error proceeds as
follows.  (We emphasize that this algorithm is quite general and could
be used with the elements of any stabilizer to find compatible Hamiltonians.)  
\begin{itemize}
\item[(1)] Expand $H$ in terms of the complete 
set of Hermitian matrices $\mu_{ijk}$ described above:
$H=\sum_{ijk}a_{ijk}\mu_{ijk}$.

\item[(2)] Determine the commutator of the general Hamiltonian $H$ and 
a generic collective error $\sum_i g_i S_i$, where the $g_i$ are
arbitrary coefficients.  
In other words, calculate $[H,\sum_i g_i S_i]$.  

\item[(3)] Find the projection of $\left[H,\sum_i g_i S_i\right]$ onto a component 
$\mu_{ijk}$ by taking the trace of the basis element $\mu_{ijk}$ with the 
result of (2).  In other words, calculate $\tr\left(\left[H,\sum_i g_i
S_i\right]\mu_{ijk}\right)$ for each $\mu_{ijk}$.  

\item[(4)] Set all of the projections equal to zero and then solve the
system of linear equations for the 
expansion coefficients $a_{ijk}$ which satisfy these relations, thereby
determining the $H$ which will commute with $\sum_i g_i S_i$.  
\end{itemize}

The results are summarized in the following 
\begin{eqnarray}
h_1 &=& \sum_{i=1}^8\mu_{0ii}, \label{eq:h1} \\ 
h_2 &=& \sum_{i=1}^8\mu_{i0i},  \label{eq:h2} \\
h_3 &=& \sum_{i=1}^8\mu_{ii0},  \label{eq:h3} \\
h_4 &=& \sum_{ijk \neq 0} f_{ijk} \mu_{ijk},  \label{eq:h4} \\
h_5 &=& \sum_{ijk \neq 0} d_{ijk} \mu_{ijk},\label{eq:h5}
\end{eqnarray}
where the $f_{ijk}$ are the structure constants and the $d_{ijk}$ are
components of the totally symmetric $d$-tensor.  (See
\ref{app:sunalg}.)  These are the only Hamiltonians, along with
combinations of these, which commute with all of the collective errors
that may act on a set of three qutrits.  The first three are
generalizations of the Heisenberg exchange operations which are known
to be universal for the three- and four-qubit DFS/NSs \cite{Kempe:00}.
The last two are, perhaps, not immediately obvious candidates for DFS
compatible Hamiltonians.  However, we will show analytically that
these two, along with the first three, are compatible with all three
qudit NSs.


\subsection{Hamiltonians Compatible with a 3-qudit DFS} 

\label{sec:dits}

We will now show that a set of Hamiltonians having the 
form of Eqs.~(\ref{eq:h1})-(\ref{eq:h5}) are compatible 
with a three-{\it qudit} DFS/NS and can be 
obtained analytically. This follows from the fact that
Eq.~(\ref{eq:larels}), the Jacobi identity Eq.~(\ref{eq:jid}), and
Jacobi-like identity 
Eq.~(\ref{eq:jlid}) hold for all SU($d$), $d\geq 3$
\cite{dgatenote1}.  

Let us consider a Hamiltonian constructed from 
the $\mu_{ijk}$ with the constituent basis elements 
$\lambda_i$ belonging to a matrix representation 
of the Lie algebra of SU$(d)$ with $d$ arbitrary:
\begin{equation}
H = \sum_{ijk}a_{ijk}\mu_{ijk}.
\end{equation}
The three qutrit case above is a special case for SU($d$) when $d=3$.  
If we consider collective errors on the DFS/NS, 
$$
S_j=\mu_{j00}+\mu_{0j0}+\mu_{00j},
$$
our condition, Eq.~(\ref{eq:HScond}) reads
$$
[S_j,H] = 0,
$$
which implies that we require 
\begin{equation}
\label{eq:cocond}
f_{ilm}a_{jkl} + f_{jlm}a_{kil} + f_{klm}a_{ijl} =0.
\end{equation}
Comparing this equation with Eqs.~(\ref{eq:jid}) and (\ref{eq:jlid}), 
we can immediately see that there are two sets of numbers $a_{ijk}$ 
which satisfy Eq.~(\ref{eq:cocond}), $f_{ijk}$ and $d_{ijk}$.  
In other words, this equation is satisfied for  $a_{ijk}=f_{ijk}$ 
and for $a_{ijk}=d_{ijk}$.  

To obtain Hamiltonians of the form 
$h_1,h_2,h_3$, we let one of the factors 
be the identity in $\mu_{ijk}$.  Let us consider one particular 
case, the case when the third factor is the identity 
$$
H = \sum_{i,j=1}^Na_{ij0}\mu_{ij0}
$$
and calculate $[H,S_j]$.  The result is 
$$
a_{il0}f_{lkj}+a_{lj0}f_{lki} = 0.  
$$
If we multiply by $d_{mki}$ and sum over $k$ and $i$, we 
get
$$
a_{il0}f_{lkj}d_{mki} = 0.
$$
Clearly, if $a_{il0}=\delta_{il}$ this equation is 
satisfied and we obtain the form Eq.~(\ref{eq:h3}).  
Similarly for  Eq.~(\ref{eq:h1}) and Eq.~(\ref{eq:h2}).  
We emphasize that the analytic proof is valid for 
all three-qudit DFS/NSs, not just qutrit states.  

Also, we have not 
proved the converse; i.e., we have not shown that these are 
the only Hamiltonians which commute with collective errors on 
three qudits.  The algorithm in the previous section 
has been implemented for 
qutrits and, in that case, these are the only Hamiltonians 
satisfying the stated conditions for DFS/NS compatibility.  
We suspect that this is also the case for qudits.


\subsection{Generalizations}  

It is important to note that several of these results are quite general.  
First, we
reiterate that the algorithm used to find the Hamiltonians for the
collective three qutrit DFS/NS can be used for any DFS/NS given a basis for the
stabilizer elements of the code space.  Second, as shown later, due to
the permutation symmetry of qudits undergoing collective errors, the
Hamiltonians we have derived are compatible with any qudit DFS/NS
undergoing collective errors.  Perhaps even more generally stated, 
the $d$-state system 
Hamiltonians with the same form as Eqs.~(\ref{eq:h1})-(\ref{eq:h5}), 
including the generalized exchange interaction, 
commute with any collective transformation on a set of qudits.
  
Furthermore, as shown in the next section, the exchange Hamiltonians can be
analytically exponentiated to provide generalized SWAP operations.
The SWAP generates the permutation group on the set of
qudits and this group commutes with the group of collective
unitary transformations.  A description of this is given in 
Ref.~\cite{Bacon/etal:07} where it is shown that one can use this
relation to produce efficient qudit circuits.


\section{Universal Computation}
\label{sec:unicomp}

We may now ask the question, are these Hamiltonians 
sufficient to perform arbitrary unitary transformations 
on NS qubits?  In other words, can we perform universal 
computation on this NS using the compatible Hamiltonians 
provided here?

In order to answer these questions, we will first note some 
important properties of the algebra of the Hamiltonians we 
have found to be DFS-compatible.  Then we attempt to 
find the corresponding unitary transformations.  


\subsection{Commutation Relations}

Now let $N$ = $d^2$ - 1 be the number of matrices in a basis 
for the Lie algebra of $SU(d)$.  
We will define the $d$-dimensional analogs of 
Eqs.~(\ref{eq:h1}) - (\ref{eq:h5}) by
\begin{eqnarray}
e_1(d) &=& \sum_{i=1}^N\mu_{0ii}, \label{eq:e1} \\
e_2(d) &=& \sum_{i=1}^N\mu_{i0i},  \label{eq:e2}\\
e_3(d) &=& \sum_{i=1}^N\mu_{ii0},  \label{eq:e3}\\
F(d)   &=& \sum_{ijk \neq 0} f_{ijk} \mu_{ijk},  \label{eq:F}\\
D(d)   &=& \sum_{ijk \neq 0} d_{ijk} \mu_{ijk}.\label{eq:D}
\end{eqnarray}
where the $\mu_{ijk}$ now represent the tensor product of 
three $d\times d$ basis matrices. It can be shown that the 
commutation of any 
two different $e_i(d)$ Hamiltonians yields the $F(d)$ Hamiltonian, 
\begin{eqnarray}
[e_1(d),e_2(d)] &=& -2iF(d), \label{eq:comm1}\\
\left[e_1(d),e_3(d)\right] &=& +2iF(d), \label{eq:comm2}\\
\left[e_2(d),e_3(d)\right] &=& -2iF(d). \label{eq:comm3}
\end{eqnarray}
One can also show that the commutation of an $e_i(d)$ 
Hamiltonian with the $F(d)$ Hamiltonian gives a combination 
of the remaining two $e_i$'s.  
\begin{eqnarray}
[e_1(d),F(d)] &=& 4i(e_2(d) - e_3(d)), \label{eq:comm4}\\
\left[e_2(d),F(d)\right] &=& 4i(e_3(d) - e_1(d)), \label{eq:comm5}\\
\left[e_3(d),F(d)\right] &=& 4i(e_1(d) - e_2(d)). \label{eq:comm6}
\end{eqnarray}

Furthermore, we have found that the $D(d)$ Hamiltonian 
commutes with the three $e_i$'s as well as $F(d)$,
\begin{equation}
[e_i(d),D(d)] = 0, \;  i = 1,2,3 \label{eq:comm7}
\end{equation}
and
\begin{equation}
[F(d),D(d)] = 0. \label{eq:comm8}
\end{equation}
This last relation, Eq.~(\ref{eq:comm8}), can be obtained by an 
explicit expansion of the two ordered products using Eq.~(\ref{eq:larels}). 
The expansion may be reduced 
to terms involving products of the form $f_{ijm}f_{klm}$ which 
can be expanded using Eq.~(\ref{eq:Mac(2.10)}). What remains can 
be reduced further with the help of Eqs.~(\ref{eq:Mac(2.17)}) and 
(\ref{eq:Mac(2.18)}) giving the stated result.  

With these results, we may now show that a sub-algebra isomorphic 
to the Lie algebra of SU(2) is generated by a combination of 
these Hamiltonians.  First, note that from 
Eqs.~(\ref{eq:comm4}) and (\ref{eq:comm5}), we may show that 
\begin{equation}
[(e_1-e_2),F] = 4i(e_1 + e_2 - 2e_3).  
\end{equation}
From Eqs.~(\ref{eq:comm1})-(\ref{eq:comm3})
\begin{equation}
 [(e_1-e_2),(e_1 + e_2 - 2e_3)] = -12iF.
\end{equation}
Therefore, the three matrices $(e_1-e_2)/2\sqrt{3},
(e_1 + e_2 - 2e_3)/6$, and $F/2\sqrt{3}$ 
form a representation of the Lie algebra of SU(2).  We will 
now use this result in the construction of the logical 
analogues of the Pauli matrices acting on the encoded qubit states.


\subsection{Unitary Transformations and Logical Operations for Qudits}

Before obtaining the logical gating operations, it is interesting 
to note that analytic expressions for the exponential of 
each one of the three $e_j$ Hamiltonians in the physical 
basis may be obtained for qudits.  There are two fortuitous properties of these 
matrices which enable us to provide such an analytic expression: 
1) the sum of 
the off-diagonal matrices commutes with the sum of the diagonal matrices and 
2) when squared, the sum of the off-diagonal matrices is 
diagonal and can easily be summed.  This can be shown to 
be true by direct computation using the set of Gell-Mann matrices 
in the case of qutrits.  \ref{app:quditgates} 
provides proof that it is also true for qudits.  
These two properties enable us to sum the series 
resulting from the exponential of the Hamiltonians 
$e_1, e_2, e_3$.  An explicit form for the unitary evolution
corresponding to these Hamiltonians is also given in  
\ref{app:quditgates}.  

Now let us define the logical ``X'' operator, $\mathbf{X}$, 
which acts on the DFS through the relation 
\begin{equation}
\label{eq:logicalX}
\bar{\mathbf{X}} = \frac{1}{2\sqrt{3}}(e_1 - e_2).
\end{equation}
We note that the overall sign of the states spanning logical one
Eq.~(\ref{eq:octet2}) are chosen so that 
the form of $\bar{\mathbf{X}}$ above resembles the expressions
appearing in Eq.~(\ref{eq:3qbXgate}) and Eq.~(\ref{eq:4qbXgate}) for 
the three- and four- qubit DFSs, respectively. It should also be
mentioned that we are describing the logical X operation 
in terms of Hamiltonians rather than unitary transformations. 
For comparison, notice that Eqs.~(\ref{eq:3qbXgate}) and
(\ref{eq:4qbXgate}) could also be written as 
$\frac{1}{2\sqrt{3}}(\vec{\sigma }_{2}\cdot \vec{\sigma }_{3} - \vec{\sigma }_{1}\cdot \vec{\sigma }_{3})$.
 
The exponentiation of $\bar{\mathbf{X}}$ leads to a time evolution given by
\begin{equation}
\label{eq:unitaryX}
U_{\bar{\mathbf{X}}} = \Bid + i\bar{\mathbf{X}}\sin(t) - \bar{\mathbf{X}}^2(1 - \cos(t)).
\end{equation}
This can be obtained in two different ways.  
One way is to use the 
prescription for the exponential of the $e_i$ as described above.  
Another one is to calculate the exponential from the results of \ref{app:xbar} and then transform back to the 
physical basis using the fact that for a similarity transformation 
$V$, and a Hamiltonian $H$,  
\begin{equation}
V\exp(-i H t)V^{-1} = \exp(-i VHV^{-1}t).  
\end{equation}
This enables one to transform between the physical, or computational 
basis states and the logical basis states as well as between the 
different sets of 
operators - the physical unitary transformations one 
would implement in experiments and the logical ones.  

Similarly, the logical ``Z'' operator, $\mathbf{Z}$, can be expanded in terms 
of the $e_i$'s by 
\begin{equation}
\label{eq:logicalZ}
\bar{\mathbf{Z}} = \frac{1}{6} (e_1 + e_2 - 2e_3). 
\end{equation}
The form of the unitary transformation resulting from the 
exponentiation of $\bar{\mathbf{Z}}$ is similar 
to that of $U_{\bar{\mathbf{X}}}$ and is given by
\begin{equation}
\label{eq:unitaryY}
U_{\bar{\mathbf{Z}}} = \Bid - i\bar{\mathbf{Z}}\sin(t) - \bar{\mathbf{Z}}^2(1 - \cos(t)).
\end{equation}

Once $\bar{\mathbf{X}}$ and $\bar{\mathbf{Z}}$ are found 
the logical $\bar{\mathbf{Y}}$ can be produced through commutation.  
This shows, with our particular choice of scaling, that the 
three matrices proportional to 
$(e_1 + e_2 - 2e_3), \; (e_2-e_1)$ and $F$ form a representation of 
the Lie algebra of SU(2) and act as the Pauli matrices on the 
DFS qubit made from three qutrits.  

A rotation in $SU(2)$ about an arbitrary axis can be obtained by three 
successive rotations.  In particular, a rotation 
about the logical $\bar{\mathbf{Y}}$ may be performed using the 
following decomposition of the logical $SU(2)$ group, using 
real (Euler) angles $\alpha, \beta,$ and $\gamma$:
\begin{equation}
U(\alpha,\beta,\gamma) = 
      \exp[-i\bar{\mathbf{Z}}\alpha]\exp[-i\bar{\mathbf{X}}\beta]\exp[-i\bar{\mathbf{Z}}\gamma].
\end{equation}
Indeed this parametrizes all of the group SU(2).  Therefore, we have shown 
that for an encoded qubit comprised of three qudits, these DFS/NS compatible operations alone can perform 
any rotation over the logical subsystems.


\subsection{Unitary Transformations and Logical Operations for 
Qutrits} 

In the logical basis, $U_{\bar{\mathbf{X}}}$ acts as the identity on 
the singlet and decuplet states (given in Eq.~(\ref{eq:decomposition}), 
and explicitly in \cite{Byrd:06}) 
throughout the entire evolution.  The action of $\bar{\mathbf{X}}$ 
on the states forming the logical 
zero Eqs.~(\ref{eq:octet1}) is such that it swaps the states for 
their logical one counterparts Eqs.~(\ref{eq:octet2}) and 
vice versa, 
\begin{eqnarray}
\label{eq:XonOctets}
\bar{\mathbf{X}}\psi_j^{8,0} &=&  \psi_j^{8,1}, \\
\bar{\mathbf{X}}\psi_j^{8,1} &=&  \psi_j^{8,0}.
\end{eqnarray}
In other words, it acts as a Pauli $X$ gate on the logical states. 
These relations can be used along with Eq.~(\ref{eq:unitaryX}) to 
calculate the action of $U_{\bar{\mathbf{X}}}$ on logical zero 
basis states,
\begin{eqnarray}
\label{eq:UXonOctet1}
U_{\bar{\mathbf{X}}}(\psi_j^{8,0}) &=& [\Bid 
+ i\bar{\mathbf{X}}\sin(t) - \bar{\mathbf{X}}^2(1 - \cos(t))]\psi_j^{8,0} \nonumber \\
&=& \psi_j^{8,0} + i\sin(t)\psi_j^{8,1} - (1 - \cos(t))\psi_j^{8,0} \nonumber \\
&=& \cos(t)\psi_j^{8,0} + i\sin(t)\psi_j^{8,1}.
\end{eqnarray}
It therefore follows that
\begin{equation}
\label{eq:UXonLogical0}
U_{\bar{\mathbf{X}}}\ket{0_L} = \cos(t)\ket{0_L} + i\sin(t)\ket{1_L},
\end{equation}
and similarly,
\begin{equation}
\label{eq:UXonLogical1}
U_{\bar{\mathbf{X}}}\ket{1_L} = \cos(t)\ket{1_L} + i\sin(t)\ket{0_L}.
\end{equation}
Now note that the logical ``Z'' operator, 
$\bar{\mathbf{Z}}$, acts as 
the identity on states in octet 1 while changing the overall sign 
of states in octet 2 and this can be used to obtain 
Eq.~(\ref{eq:logicalZ}) directly for qutrits using the 
explicit expressions.  The unitary may then also be 
obtained directly from $\bar{\mathbf{Z}}$.  
The transformation of logical zero is given by 
\begin{eqnarray}
\label{eq:UZonLogical0}
U_{\bar{\mathbf{Z}}}\ket{0_L} &=& U_{\bar{\mathbf{Z}}}\sum_{j}\alpha_{j}\psi_j^{8,0} 
= [\Bid - i\bar{\mathbf{Z}}\sin(t) - 
          \bar{\mathbf{Z}}^2(1 - \cos(t))]\sum_{j}\alpha_{j}\psi_j^{8,0} \nonumber \\
&=& \sum_{j}\alpha_{j}\psi_j^{8,0} - i\sin(t)\sum_{j}\alpha_{j}\psi_j^{8,0} 
 - (1 - \cos(t))\sum_{j}\alpha_{j}\psi_j^{8,0} \nonumber \\
&=& \ket{0_L}\exp(-it),
\end{eqnarray}
while logical one transforms unitarily by
\begin{equation}
\label{eq:UZonLogical1}
U_{\bar{\mathbf{Z}}}\ket{1_L} = \ket{1_L}\exp(+it).
\end{equation}
Again, the decuplet states are left unchanged by the action of 
$U_{\bar{\mathbf{Z}}}$. This implies, along with the invariance 
of the singlet state, that this gate set is canonical in the sense 
described in \cite{Byrd/etal:05} which is important for applications 
of decoupling pulses to eliminate leakage and protect the information 
\cite{Wu/etal:02,Byrd/etal:05}.


\subsection{SWAP Operation}
\label{sec:swap}

The exchange, or SWAP operation between qudits $p$ and 
$q$ can be achieved by allowing the appropriate 
DFS/NS compatible Hamiltonian $e_m(d)$, $(p\neq m \neq q)$ to act 
between the two for a specific amount of time. To show this, 
let us write 
\beq
\sum_{s=1}^{d^2-1}\lambda_s \otimes \lambda_s =\sum_i 
\lambda_i \otimes \lambda_i + \sum_j 
\lambda_j \otimes \lambda_j,
\eeq 
where the $\lambda_i$ ($\lambda_j$) represent the diagonal (off-diagonal) 
components of the traceless, Hermitian basis $\{\lambda_s\}$ normalized 
such that Tr$(\lambda_s\lambda_{s^\prime}) =
2\delta_{ss^\prime}$. ({\it In this section only, we use} $i$ {\it for
indices on diagonal elements of the algebra and} $j$ {\it for elements of
the algebra which have no nonzero diagonal elements.})  Since 
$\sum_i \lambda_i \otimes \lambda_i$ commutes with $\sum_j 
\lambda_j \otimes \lambda_j$ (see \ref{app:dnodcomm}) we 
may express the exponential of 
$\sum_{s=1}^{d^2-1}\lambda_s \otimes \lambda_s$ as
\beq
\!\!\!\!\!\!\!\!\!\!
\label{eq:swap}
\exp\left[-it\sum_{s=1}^{d^2-1}\lambda_s \otimes \lambda_s\right]=
\exp\left[-it\sum_i \lambda_i \otimes \lambda_i\right]\exp\left[-it\sum_j 
\lambda_j \otimes \lambda_j\right].
\eeq

The off-diagonal members of this basis may be written either as 
$\dmat{k}{l}+\dmat{l}{k}$ or $i\dmat{k}{l}-i\dmat{l}{k}$ for 
$k \neq l$ (for notational simplicity we use the two-qudit 
computational basis $(\{\ket{11},\ket{12},\;.\;.\;.\;,\ket{dd}\})$, 
so that the off-diagonal contribution appearing 
in this last equation may be written as 
\beq
\label{eq:offd}
\exp\left[-it\sum_j \lambda_j \otimes \lambda_j\right] = 
\exp\left[-it\sum_{k<l}M_{k,l}\right],
\eeq
 where
\bea
M_{k,l} &\equiv& (\dmat{k}{l}+\dmat{l}{k}) \otimes (\dmat{k}{l}+\dmat{l}{k})
\nonumber \\ &+& 
(i\dmat{k}{l}-i\dmat{l}{k}) \otimes (i\dmat{k}{l}-i\dmat{l}{k}) \nonumber \\
&=& 2\dmat{k}{l} \otimes \dmat{l}{k} + 2\dmat{l}{k} \otimes \dmat{k}{l}.
\eea
For $SU(d)$ the number of $M_{k,l}$ appearing in the summation of 
Eq.~(\ref{eq:offd}) 
is $(d^2 - d)/2$. Also, each distinct $M_{k,l}$ can be seen to commute with 
the others since $
(\dmat{k}{l} \otimes \dmat{l}{k} + \dmat{l}{k} \otimes \dmat{k}{l})
\times (\dmat{p}{q} \otimes \dmat{q}{p} + \dmat{q}{p} \otimes \dmat{p}{q})
= 0$ when $p \neq l \neq q$ and $p \neq k \neq q$. This allows us to write 
\beq
\label{eq:jj}
\exp\left[-it\sum_j \lambda_j \otimes \lambda_j\right] = 
\exp\left[-itM_{1,2}\right]\;.\; .\; .\; \exp\left[-itM_{d-1,d}\right].
\eeq

When squared, each $M_{k,l}$ is diagonal and given by
$
M_{k,l}^2 = 4\dmat{k}{k} \otimes \dmat{l}{l} + 4\dmat{l}{l} \otimes \dmat{k}{k},
$
which implies that $M_{k,l}^3 = 4M_{k,l}$, etc. Now, if we choose to define 
\beq
\label{eq:qkl}
Q_{k,l} \equiv \frac{M_{k,l}}{2} = \dmat{k}{l} \otimes \dmat{l}{k} + \dmat{l}{k} \otimes \dmat{k}{l}, \;\; l\neq k
\eeq
and
\beq
\label{eq:rkl}
R_{k,l} \equiv \frac{M_{k,l}^2}{4} = \dmat{k}{k} \otimes \dmat{l}{l} + \dmat{l}{l} \otimes \dmat{k}{k}, \;\; l \neq k
\eeq
we may express the exponential of $M_{k,l}$ as
\beq
\label{eq:expm}
U_{k,l}(t) =  \exp[-itM_{k,l}] = \Bid -iQ_{k,l}\sin(2t) + R_{k,l}(\cos(2t)-1).
\eeq
If we now let $t = \pi/4$ we find that
\beq
U_{k,l}(t=\pi/4) = \Bid - iQ_{k,l} - R_{k,l}.
\eeq
Using the definition provided in Eq.~(\ref{eq:qkl}) we see that the only 
two-qudit computational basis states $\ket{\alpha\beta}$ which survive the action of a 
particular $Q_{k,l}$ are $\ket{lk}$ and $\ket{kl}$, i.e.,
\beq
Q_{k,l}\ket{\alpha\beta} = \delta_{\alpha,l}\delta_{\beta,k}\ket{kl} + \delta_{\alpha,k}\delta_{\beta,l}\ket{lk}.
\eeq 
 Also, using Eq.~(\ref{eq:rkl}), we find that 
\beq
(\Bid -R_{k,l})\ket{\alpha\beta} = \ket{\alpha\beta} - \delta_{\alpha,l}\delta_{\beta,k}\ket{\alpha\beta} 
-\delta_{\alpha,k}\delta_{\beta,l}\ket{\alpha\beta}.
\eeq
Therefore, at $t=\pi/4$ the two-qudit computational basis states $\ket{\alpha\beta}$ evolve to 
\beq
U_{k,l}(\pi/4)\ket{\alpha\beta}=\ket{\alpha\beta} - \delta_{\alpha,l}\delta_{\beta,k}(i\ket{kl}+\ket{\alpha\beta})
-\delta_{\alpha,k}\delta_{\beta,l}(i\ket{lk}+\ket{\alpha\beta}).
\eeq

In other words, $U_{k,l}(\pi/4)$ acts as the identity on all product states $\ket{\alpha\beta}$ 
except $\ket{kl}$ and $\ket{lk}$. On these states $U_{k,l}(\pi/4)$ exchanges $\ket{kl}$ ($\ket{lk}$) for 
$\ket{lk}$ ($\ket{kl}$) along with a phase shift of $\exp(-i\pi/2)$. The action of 
$\exp[-it\sum_{s=1}^{d^2-1}\lambda_s \otimes \lambda_s]$ at $t=\pi/4$ on a given product state 
$\ket{\alpha\beta}$, $\alpha,\beta = 1,2,\;.\;.\;.\;d$ can now be calculated using 
Eqs.~(\ref{eq:swap}) and (\ref{eq:jj}).

\begin{equation}
\!\!\!\!\!\!\!\!\!\!\!\!\!\!\!\!\!\!\!\!\!\!\!\!\!\!\!\!\!\!\!
\label{eq:swap2}
\exp\left[-i(\pi/4)\sum_{s=1}^{d^2-1}\lambda_s \otimes \lambda_s\right]
\ket{\alpha\beta} 
= \left\{\begin{array}{ll} 
 \;\;\;\;\;\exp(-i(\pi/4)\sum_i \lambda_i \otimes \lambda_i)\ket{\alpha\alpha} ,&  
\mbox{if } \; \alpha = \beta, \\   
-i\exp(-i(\pi/4)\sum_i \lambda_i \otimes \lambda_i)\ket{\beta\alpha},  & 
\mbox{if} \;\; \alpha \neq \beta. 
\end{array}\right.
\end{equation}
In order to determine the diagonal elements of 
$\exp(-it\sum_i \lambda_i \otimes \lambda_i)$ we must first find 
the coefficients $\xi_{m,n}$ which satisfy the relation 
\beq
\label{eq:daigcoeff}
\sum_i \lambda_i \otimes \lambda_i = 
\sum_{m,n=1}^{d}\xi_{m,n}\dmat{m}{m} \otimes \dmat{n}{n}.
\eeq

In the following analysis we will take a Gell-Mann basis 
for $SU(d)$, in which there are a total of $d-1$ diagonal elements. 
All of these diagonal matrices can be constructed using 
Eq.~(\ref{eq:diagmat}). For $SU(3)$ the two diagonal 
elements are given by $\lambda_3= \dmat{1}{1} - \dmat{2}{2}$ and 
$\lambda_8 = (\dmat{1}{1} + \dmat{2}{2} -2\dmat{3}{3})/\sqrt{3}$. Two of 
the three diagonal elements of $SU(4)$ can be formed by respectively 
placing $\lambda_3$ and $\lambda_8$ in the upper left block of a 
$4 \times 4$ matrix with zero entries on the fourth row and column. 
The remaining 
diagonal element is then given by Eq.~(\ref{eq:diagmat}) with $d=4$, 
i.e., 
$\lambda_{15} = (\dmat{1}{1} + \dmat{2}{2} + \dmat{3}{3} 
- 3\dmat{4}{4})/\sqrt{6}$. This procedure can be used to construct 
the set of diagonal matrices corresponding to $SU(d)$ for all 
$d \geq 3$; one simply places the diagonal elements associated with 
$SU(d-1)$ into the upper left block of a $d \times d$ matrix with 
zeros along the $d^{th}$ row and column. Eq.~(\ref{eq:diagmat}) can 
be used to obtain the remaining element.

As we proceed to determine the expansion coefficients $\xi_{m,n}$ 
appearing in Eq.~(\ref{eq:daigcoeff}) let us now write the sum of 
off-diagonal matrices explicitly as 
\bea
\label{eq:odexp}
\!\!\!\!\!\!\!\!\!\!\!\!\!\!\!\!\!
\sum_i \lambda_i \otimes \lambda_i &=& (\dmat{1}{1} - \dmat{2}{2}) \otimes 
(\dmat{1}{1} - \dmat{2}{2}) + \; . \; . \; .
\nonumber \\
&+& \frac{2}{d(d-1)}(\dmat{1}{1} + \;.\;.\;.\; + \dmat{d-1}{d-1} 
- (d-1)\dmat{d}{d}) 
\nonumber \\
&& \;\;\;\;\;\;\;\;\; \otimes
(\dmat{1}{1} + \;.\;.\;.\; + \dmat{d-1}{d-1} - (d-1)\dmat{d}{d}).
\eea
If we consider the case when $m=n$ we find that the coefficient 
$\xi_{m,m}$ attached to the matrix element 
$\dmat{m}{m} \otimes \dmat{m}{m}$ obeys the relation
\begin{equation}
\label{eq:swap3}
\xi_{m,m} 
= \left\{\begin{array}{ll}  
\frac{2(d-1)}{d},  & 
\mbox{if} \;\; m = d, \\
\frac{2(m-1)}{m} + 2\sum_{\mu = m+1}^d \frac{1}{\mu(\mu-1)},  &
\mbox{if} \;\;  m \neq d. 
\end{array}\right.
\end{equation}
Let us now evaluate the series which appears in the expression 
for $\xi_{m,m}$ when $m \neq d$.
\bea
\label{eq:series}
\!\!\!\!\!\!\!\!\!\!\!\!\!\!\!\!\!\!\!\!\!\!\!\!\!\!\!\!\!\!\!\!\!\!
\sum_{\mu = m+1}^d \frac{1}{\mu(\mu-1)}&=& 
\frac{1}{m(m+1)} + \frac{1}{(m+1)(m+2)} + \;.\;.\;.\;
+\frac{1}{(d-2)(d-1)}+\frac{1}{(d-1)d} 
\nonumber \\
&=& \frac{1}{m}-\frac{1}{m+1}+\frac{1}{m+1} -\;.\;.\;.\;
-\frac{1}{d-1}+\frac{1}{d-1}-\frac{1}{d} \nonumber \\
&=& \frac{1}{m}-\frac{1}{d}= \frac{d-m}{md}
\eea
Therefore, when $m \neq d$ we find that the coefficient $\xi_{m,m}$ 
can be expressed as 
\beq
\frac{2(m-1)}{m} + 2\sum_{\mu = m+1}^d \frac{1}{\mu(\mu-1)}
= \frac{2(m-1)}{m} + \frac{2(d-m)}{md}= \frac{2(d-1)}{d},
\eeq
so that 
\beq
\label{eq:allmm}
\xi_{m,m}=\frac{2(d-1)}{d}, \;\;\; \mbox{for} \; m=1,2,\;.\;.\;.\;,d.
\eeq

Now suppose that $m \neq n$. Examination of the expansion given by Eq.~(\ref{eq:odexp}) 
reveals the symmetry of all terms $\dmat{m}{m} \otimes \dmat{n}{n}$ under 
the interchange of indices $m$ and $n$. Since the relation $\xi_{m,n} =\xi_{n,m}$ 
then follows, we may assume that $n > m$ in the proceeding discussion without 
loss of generality. Given this assumption, we find that the coefficient $\xi_{m,n}$ 
appearing in Eq.~(\ref{eq:daigcoeff}) satisfies the equation
\begin{equation}
\xi_{m,n} 
= \left\{\begin{array}{ll}  
-\frac{2}{d},  & 
\mbox{if} \;\; n = d, \\
-\frac{2}{n} + 2\sum_{\mu = n+1}^d \frac{1}{\mu(\mu-1)},  &
\mbox{if} \;\;  n \neq d. 
\end{array}\right.
\end{equation}
However, this relation for $\xi_{m,n}$ when $n \neq d$ may be simplified in view of Eq.~(\ref{eq:series}),
\beq
-\frac{2}{n} + 2\sum_{\mu = n+1}^d \frac{1}{\mu(\mu-1)} = -\frac{2}{n} + \frac{2(d-n)}{nd} = -\frac{2}{d}.
\eeq
And so we find that all of the coefficients $\xi_{m,n}$ $(m \neq n)$ are equal and given by
\beq
\label{eq:allmn}
\xi_{m,n}=-\frac{2}{d}, \;\;\; \mbox{for} \;\; m,n=1,2,\;.\;.\;.\;,d \;\;\; \mbox{and} \;\; m \neq n.
\eeq
The exponential of $\sum_i \lambda_i \otimes \lambda_i = 
\sum_{m,n=1}^{d}\xi_{m,n}\dmat{m}{m} \otimes \dmat{n}{n}$ can now be evaluated,
\bea
\label{eq:expod2}
\exp\left[-it\sum_i \lambda_i \otimes \lambda_i\right] &=&
\sum_{m,n=1}^{d}\exp(-it\xi_{m,n})\dmat{m}{m} \otimes \dmat{n}{n} \nonumber \\
&=& \sum_{m}\exp(-(2it(d-1))/d)\dmat{m}{m} \otimes \dmat{m}{m} \nonumber \\
&+& \sum_{m \neq n}\exp(2it/d)\dmat{m}{m} \otimes \dmat{n}{n}.
\eea

Therefore, the unitary transformation which corresponds to the exponentiation of 
$\sum_{s=1}^{d^2-1}\lambda_s \otimes \lambda_s$ can be expressed using 
Eqs.~(\ref{eq:swap}), (\ref{eq:jj}), (\ref{eq:expm}), and (\ref{eq:expod2}). In particular, the 
two-qudit computational basis states $\ket{\alpha\beta}$, $\alpha,\beta = 1,2,\;.\;.\;.\;d$ 
evolve at $t = \pi/4$ to   
\begin{equation}
\!\!\!\!\!\!\!\!\!\!\!\!\!\!\!\!\!\!\!\!\!\!\!\!\!\!\!\!\!\!\!
\label{eq:swap4}
\exp\left[-i(\pi/4)\sum_{s=1}^{d^2-1}\lambda_s \otimes \lambda_s\right]
\ket{\alpha\beta} 
= \left\{\begin{array}{ll} 
 \exp(-({\pi}i(d-1))/2d)\ket{\alpha\alpha} ,&  
\mbox{if } \; \alpha = \beta, \\   
-i\exp({\pi}i/2d)\ket{\beta\alpha},  & 
\mbox{if} \;\; \alpha \neq \beta. 
\end{array}\right.
\end{equation}
However, since $\exp(-({\pi}i(d-1))/2d) = \exp(-{\pi}i/2)\exp({\pi}i/2d) = -i\exp({\pi}i/2d)$, we 
have
\beq
\!\!\!\!\!\!\!\!\!\!\!\!\!\!\!\!\!\!\!\!\!\!\!\!\!\!\!\!\!\!
\label{eq:swap5}
\exp\left[-i(\pi/4)\sum_{s=1}^{d^2-1}\lambda_s \otimes \lambda_s\right]
\ket{\alpha\beta} = -i\exp({\pi}i/2d)\ket{\beta\alpha} \;\;\; \mbox{for} \;\; \alpha,\beta = 1,2,\;.\;.\;.\;,d.
\eeq
Although the exponential of $\sum_{s=1}^{d^2-1}\lambda_s \otimes \lambda_s$ induces a phase 
shift on all product states $\ket{\alpha\beta}$, the shift is the same for all 
initial states as they evolve to $t=\pi/4$. This unitary transformation thus acts as the SWAP operation 
at $t=\pi/4$ between any two $d$-state systems.


\section{Hamiltonians Compatible With an n-Qudit DFS} 

\label{sec:ndits}

In this section we show that the Hamiltonians given by 
Eqs.~(\ref{eq:e1})-(\ref{eq:D}) remain compatible with a collective DFS/NS 
encoding of $n$ physical qudits into a set of logical quDits, $d\neq D$. 
Since, in principle,  a logical quDit of 
dimension $D$ can be formed by such an encoding, the Hamiltonians 
considered here have the ability to generate compatible transformations 
on an encoded quDit state of arbitrary dimension. To see this, let 
us first consider the exchange operations $e_i(d)$ which were previously 
shown to commute with the collective errors, i.e., 
\bea
\label{eq:ndit1}
\left[e_1(d),S_j\right] &=& 
\left[\Bid \otimes \lambda_i \otimes \lambda_i \:,\:
\lambda_j \otimes \Bid \otimes \Bid + 
\Bid \otimes \lambda_j \otimes \Bid +
\Bid \otimes \Bid \otimes \lambda_j \right] \nonumber \\
&=& \Bid \otimes (\lambda_i \lambda_j - \lambda_j \lambda_i) \otimes \lambda_i 
+ \Bid \otimes \lambda_i \otimes (\lambda_i \lambda_j - \lambda_j \lambda_i)
\nonumber \\
&=& 2if_{ijk}(\Bid \otimes \lambda_k \otimes \lambda_i 
- \Bid \otimes \lambda_k \otimes \lambda_i) \nonumber \\
&=& 0.
\eea
(Recall that the structure constants $f_{ijk}$ are totally 
antisymmetric and that the generators of $SU(d)$ obey 
the commutation relations $[\lambda_i,\lambda_j] =
2if_{ijk}\lambda_k$. Again, summation over repeated indices is implied.) 
Notice how the term $\left[\lambda_i \otimes \lambda_i \otimes \Bid \:,\: 
\Bid \otimes \Bid \otimes \lambda_j \right]$ in the commutator of 
$\left[e_1(d),S_j\right]$ can be immediately neglected while the 
nontrivial contributions are proportional to 
$\left[\lambda_i,\lambda_j \right] \otimes \lambda_i + \lambda_i \otimes 
\left[\lambda_i,\lambda_j \right]$. The commutators 
of $e_2(d)$ and $e_3(d)$ with the collective errors 
also share these properties,
\bea
\label{eq:ndit2}
\left[e_2(d),S_j\right] &=& 
\left[\lambda_i \otimes \Bid \otimes \lambda_i \:,\:
\lambda_j \otimes \Bid \otimes \Bid + 
\Bid \otimes \lambda_j \otimes \Bid +
\Bid \otimes \Bid \otimes \lambda_j \right] \nonumber \\
&=& \left[\lambda_i,\lambda_j \right] \otimes \Bid \otimes \lambda_i 
+ \lambda_i \otimes \Bid \otimes \left[\lambda_i,\lambda_j \right]  
\nonumber \\
&=& 2if_{ijk}(\lambda_k \otimes \Bid \otimes \lambda_i 
- \lambda_k \otimes \Bid \otimes \lambda_i) \nonumber \\
&=& 0,
\eea
and similarly for $\left[e_3(d),S_j\right]$. We now allow 
the system to be composed of $n$ physical qudits.  Define
\beq
e_{pq} = \sum_i \lambda^{(p)}_i\lambda^{(q)}_i, 
\eeq
where the superscripts identify qudits $p$ and $q$, all other
qudits are acted upon by identity operators, and we have written the
sum explicitly for emphasis.  
We can now show that all of the generalized exchange Hamiltonians 
$e_{pq}$ for this larger system 
also commute with the collective errors $S_{\alpha}$. To see this,
note that the 
collective errors acting on the $n$ qudits are again defined as 
\beq
S_{\alpha} \equiv \sum_{r=1}^{n}\lambda_{\alpha}^{(r)}, \:\:\: \mbox{for} 
\:\:\: \alpha = 1, 2, ..., d^2-1.
\eeq
The only nontrivial contributions appearing in 
the calculation of 
$\left[e_{pq},S_{\alpha} \right]$ 
have the form
\bea 
\left[\lambda_i,\lambda_j \right]^{(p)}  \lambda_i^{(q)}
 +  \lambda_i^{(p)}
 \left[\lambda_i,\lambda_j \right]^{(q)} 
\eea
with the identity lying at all other positions. Since these terms 
may be rewritten as
\bea 
2if_{ijk}(\lambda_{k}^{(p)}  \lambda_i^{(q)} 
+  \lambda_i^{(p)}
 \lambda_{k}^{(q)}), 
\eea  
or
\bea 
2if_{ijk}(\lambda_{k}^{(p)}  \lambda_i^{(q)}
 - \lambda_{k}^{(p)} \lambda_{i}^{(q)}) = 0,
\eea  
we see that the exchange Hamiltonians are also compatible with a DFS/NS 
that is supported by $n$ physical qudits.

This generalization to $n$ qudits is also applicable to the
Hamiltonian $F(d)$.
Recall the commutator of $F(d)$ with a three particle collective error. 
In this case we have
\bea \!\!\!\!\!\!\!\!\!\!\!\!\!\!\!
\left[F(d),S_l\right] &=& \left[f_{ijk} \lambda_i \otimes \lambda_j \otimes 
\lambda_k \:,\: \lambda_l \otimes \Bid \otimes \Bid 
+ \Bid \otimes \lambda_l \otimes \Bid + \Bid \otimes \Bid \otimes \lambda_l 
\right] \nonumber \\
&=& f_{ijk}(\left[\lambda_i,\lambda_l\right] \otimes \lambda_j 
\otimes \lambda_k + \lambda_i \otimes \left[\lambda_j,\lambda_l\right] \otimes 
\lambda_k + \lambda_i \otimes \lambda_j \otimes 
\left[\lambda_k,\lambda_l\right]) \nonumber \\
&=& 2if_{ijk}(f_{ilm}\lambda_m \otimes \lambda_j \otimes \lambda_k + 
f_{jlm}\lambda_i \otimes \lambda_m \otimes \lambda_k +
f_{klm}\lambda_i \otimes \lambda_j \otimes \lambda_m). \nonumber \\
\eea 
This can reduced using Eq.~\ref{eq:Mac(2.10)}. After some relabeling 
we obtain 
\bea 
\left[F(d),S_l\right] &=&
2i(d_{jlk}d_{kmi} - d_{jlk}d_{kmi} 
+ d_{jik}d_{kml} \nonumber \\
&-& d_{jik}d_{kml}
+ d_{jmk}d_{kli} - d_{jmk}d_{kli})\: \lambda_i \otimes \lambda_j \otimes 
\lambda_m \nonumber \\
&=&0.
\eea
Now suppose that we have a system of $n$ qudits with the interaction 
coupling particles $p$, $q$, and $r$, then 
\bea
\left[F(d),S_l\right] &=& \left[f_{ijk} \lambda_i^{(p)} 
\lambda_j^{(q)}  \lambda_k^{(r)} \:,\: \lambda_l \otimes 
... \otimes \Bid + ... + \Bid \otimes ... \otimes \lambda_l\right].
\eea
Again, all of the terms in this expansion that do not contain 
ordinary products of two $\lambda$'s will cancel trivially, those which 
remain can be expressed as 
\bea \!\!\!\!\!\!\!\!\!\!\!\!\!\!\!\!\!\!\!\!\!\!
f_{ijk}(\left[\lambda_i,\lambda_l\right]^{(p)} \lambda_j^{(q)} 
\lambda_k^{(r)} + \lambda_i^{(p)}  
\left[\lambda_j,\lambda_l\right]^{(q)} \lambda_k^{(r)} + 
\lambda_i^{(p)} \lambda_j^{(q)}  
\left[\lambda_k,\lambda_l\right]^{(r)}), 
\eea  
which becomes
\bea 
\left[F(d),S_l\right] &=&
2i(d_{jlk}d_{kmi} - d_{jlk}d_{kmi} 
+ d_{jik}d_{kml} \nonumber \\
&-& d_{jik}d_{kml} + d_{jmk}d_{kli} - d_{jmk}d_{kli})\: 
\lambda_i^{(p)} \lambda_j^{(q)} \lambda_m^{(r)} \nonumber \\
&=&0.
\eea

The $D(d)$ Hamiltonian can also be shown in a similar way to commute 
with the collective errors acting on a system of $n$ qudits. Therefore, 
these DFS/NS compatible operators can be used to manipulate 
encoded quDits of arbitrary dimension as well.


\section{Conclusions}
\label{sec:concl}

We have presented a set of Hamiltonians that are 
compatible with an $n$-qudit DFS/NS ($d$ and $n$ arbitrary) 
encoding which protects against collective noise, i.e., any noise 
which affects the qudits in the same way.  We have also 
provided a set of unitary transformations which can be used to perform 
any rotation on a logical qubit that is represented by the two degenerate 
representations in the product space of three qudits.  
The generalized exchange operations are analogs of the Heisenberg exchange 
transformation which is known to be universal for various 
systems of qubits and we have shown that they, as well as the logical
gating operations, can be exponentiated analytically. In this context,
we note that the three-qudit DFS/NS is 
similar in structure to the three qubit DFS/NS.  

Our analysis has focused on Hamiltonians which commute with 
elements of the stabilizer and the unitary transformations 
arising from the exponentiation of the generalized 
exchange operation for qudits.  For three qutrits we have found 
the complete set of Hamiltonians satisfying this commutation condition.  
However, these results could be extended by using the less restrictive 
requirement provided in Eq.~(\ref{eq:uscond}).  

In order for these Hamiltonians to
enable universal quantum computation they must also be able to
generate entanglement between two encoded qubits. The CNOT gate 
acting on a subspace of the three-qubit DF-subsystem was provided in
Ref.~\cite{DiVincenzo:00a} using a circuit of 19 exchange
interactions. Since the states considered there also appear in the
three-qutrit DFS/NS as $\psi_2^{8,0}$ and $\psi_2^{8,1}$ we find that
the exchange interaction alone can implement universal quantum computing 
over the three-qutrit DFS/NS as well. Furthermore, since the tableau
for all three-${\it{qudit}}$ DFSs have an identical structure, those
same states must also appear in the expansions of the logical states
for $d\geq 3$. Therefore, the Hamiltonians derived here are sufficient
to perform universal quantum computation using the three qudit DFS/NS. 

Finally we note that our algorithm for the determination of 
compatible Hamiltonians is also applicable to DFS/NSs which are not
collective.  
In addition, the generalized exchange interaction is compatible with any
collective DFS/NS and can be analytically exponentiated to produce the 
corresponding unitary transformations, including SWAP gates, for
qudits encoding quDits.


\section*{Acknowledgments}

The authors thank Zhao-Ming Wang and Lian-Ao Wu for several helpful
comments/discussions.  
This material is based upon work supported by the National Science 
Foundation under Grant No. 0545798.


\verb``\appendix

\section{The Algebra of $SU(d)$}



\label{app:sunalg}

We have chosen the following convention for the normalization of 
the algebra of Hermitian matrices which represent generators of $SU(d)$.  
\begin{equation}
\tr(\lambda_i\lambda_j) = 2\delta_{ij}.  
\end{equation}

The commutation and anti-commutation relations of the matrices 
representing the basis for the Lie algebra can be summarized 
by the following equation:
\begin{equation}
\label{eq:larels}
\lambda_i \lambda_j = \frac{2}{d}\delta_{ij} + if _{ijk} \lambda_k 
                      + d_{ijk}\lambda_k,
\end{equation}
where here, and throughout this appendix, a sum over repeated 
indices is understood.  

As with any Lie algebra we have the Jacobi identity:
\begin{equation}
\label{eq:jid}
f_{ilm}f_{jkl} + f_{jlm}f_{kil} + f_{klm}f_{ijl} =0.
\end{equation}
There is also a Jacobi-like identity,
\begin{equation}
\label{eq:jlid}
f_{ilm}d_{jkl} + f_{jlm}d_{kil} + f_{klm}d_{ijl} =0,
\end{equation}
which was given by Macfarlane, et al. \cite{Macfarlane}. 

The following identities, also provided in \cite{Macfarlane}, are useful in 
the derivation of the commutation relation given by Eq.~(\ref{eq:comm8}),
\begin{eqnarray}
d_{iik} &=& 0, \label{eq:Mac(2.7)}\\
d_{ijk}f_{ljk} &=& 0,  \label{eq:Mac(2.14)}\\
f_{ijk}f_{ljk} &=& d\delta_{il},  \label{eq:Mac(2.12)}\\
d_{ijk}d_{ljk} &=& \frac{d^2 - 4}{d}\delta_{il},  \label{eq:Mac(2.13)}
\end{eqnarray}
and
\begin{equation}
f_{ijm}f_{klm} = \frac{2}{d}(\delta_{ik}\delta_{jl} - \delta_{il}\delta_{jk}) 
                  + (d_{ikm}d_{jlm} - d_{jkm}d_{ilm}) \label{eq:Mac(2.10)}
\end{equation}
 and finally
\begin{eqnarray}
f_{piq}f_{qjr}f_{rkp} &=& -\left(\frac{d}{2}\right)f_{ijk}, \label{eq:Mac(2.15)}\\
d_{piq}f_{qjr}f_{rkp} &=& -\left(\frac{d}{2}\right)d_{ijk}, \label{eq:Mac(2.16)}\\
d_{piq}d_{qjr}f_{rkp} &=& \left(\frac{d^2 - 4}{2d}\right)f_{ijk}, \label{eq:Mac(2.17)}\\
d_{piq}d_{qjr}d_{rkp} &=& \left(\frac{d^2 - 12}{2d}\right)d_{ijk} \label{eq:Mac(2.18)}.
\end{eqnarray} 
The proofs of these are fairly straight-forward and are omitted.


\section{Analytic Expressions for Unitary Transformations}

\label{app:quditgates}

It was stated in \ref{sec:swap} that the sum of all terms having the form 
$\lambda_i \otimes \lambda_i$ can be shown to commute with the sum 
of all terms $\lambda_j \otimes \lambda_j$, where the $\lambda_i$ $(\lambda_j)$ represent 
the diagonal (off-diagonal) basis elements of $SU(d)$. This appendix provides 
proof of this statement and gives an alternative description of the time 
evolution of the sum of off-diagonals.  


\subsection{Sum of Diagonals and Sum of Off-diagonals Commute}

\label{app:dnodcomm}

The objective of this appendix is to show that 
\begin{equation}
\left[\sum_i \mu_{ii0}, \sum_j\mu_{jj0}\right] = 0,
\end{equation}
where all $\lambda_i$ are diagonal, all 
$\lambda_j$ are off-diagonal, and $\mu_{ijk} = 
\lambda_i\otimes \lambda_j \otimes \lambda_k$.  It will be convenient to 
let $i\in I$, where $I$ represents a subset of all numbers in the 
set $1,2,..., d^2-1$ which correspond to matrices $\lambda_i$ 
which are diagonal 
and similarly $j\in J$, with $J$ corresponding to indices with 
$\lambda_j$ being off-diagonal matrices.  

First we define 
\begin{equation}
C_{od} = \sum_{ij}\left[\lambda_i\otimes\lambda_i,\lambda_j\otimes\lambda_j\right],
\end{equation}
and then proceed to show that $C_{od} =0$.  Here again the 
convention is that repeated indices are to be summed.  Using the identities 
provided in \ref{app:sunalg} we may rewrite this last equation as 
\begin{eqnarray}
C_{od} &=& -2f_{ijk}f_{ijl}\lambda_k\otimes\lambda_l 
           +2if_{ijk}d_{ijl} \lambda_k\otimes \lambda_l  \nonumber \\
       && +2f_{ijk}f_{ijl} \lambda_l\otimes \lambda_k  
          -2if_{ijk}d_{ijl} \lambda_l\otimes \lambda_k \nonumber \\
       &=& 2i(f_{ijk}d_{ijl} - f_{ijl}d_{ijk})\lambda_k\otimes \lambda_l, 
\end{eqnarray}
where the first equality follows from Eq.~(\ref{eq:larels}) and the 
second from relabeling.  Now, since 
$f_{imk} =0$ for both indices $i,m \in I$ we may establish the following 
equality
\begin{equation}
\sum_{i,j}f_{ijk}d_{ijl} = \sum_{i,m}f_{imk}d_{iml},
\end{equation}
where $m = 1,2,...,d^2-1$. Since the index $i$ now serves to distinguish the diagonal 
elements, the Jacobi-like identity, Eq.~(\ref{eq:jlid}), 
can now be expressed as
\begin{equation}
2f_{imk}d_{iml} +f_{lmk}d_{iim} = 0.  
\end{equation}
This implies that if 
\begin{equation}
\label{eq:simpcond}
\sum_i d_{iil} = 0,
\end{equation}
then $C_{od} = 0$ and the result follows.  
(Note that this is certainly true when $i$ takes all values from 
1 to $d^2-1$.  Here, the $i$ are in $I$ so it may not be obvious.)  
To show this equation is satisfied for all $d$, we first state that 
it is true by direct computation 
for the matrix representation of the Lie algebra of SU(3) using the 
Gell-Mann basis and also for SU(4) using the Gell-Mann-like basis 
for SU(4).  (It is trivially true for SU(1) and SU(2) since all 
$d_{ijk}=0$.) The only nonzero $d_{iil}$ for $i \in I$ for SU(3) are
$$
d_{338} = 1/\sqrt{3}, \;\; \mbox{and} \;\; d_{888}= -1/\sqrt{3},
$$
and for SU(4) they are 
\bea
\;\;\;\;\;\;\;\;\;\;\; d_{3,3,8} = 1/\sqrt{3}, \;\; d_{8,8,8}= -1/\sqrt{3}, \;\; d_{3,3,15} = 1/\sqrt{6},
       \nonumber \\
  \;\;\;\;\;\;\;\;\;\;\;\;\;\;\;\;\;\;\; d_{8,8,15} =1/\sqrt{6}, \;\; 
d_{15,15,15} = -2/\sqrt{6}.\;\;\;\;\;\;\;\;\;\;
\eea
Clearly, in both cases, $\sum_i d_{iil}=0$.  
We now proceed by induction, showing Eq.~(\ref{eq:simpcond}) to be true 
for $d$ while assuming it is valid for $d-1$. 
We also 
note that this assumption and proof is motivated quite well by the 
$d_{iik}$ for SU(3) and SU(4).  

By convention, we take a Gell-Mann basis for SU(d) which has its diagonal 
elements of the form
\begin{equation}
\label{eq:diagmat}
\lambda_{d^2-1} = \sqrt{\frac{2}{d(d-1)}}\; \left(\begin{array}{ccccc}
                             1 & 0 & 0 & \cdots  & 0 \\
			     0 & 1 & 0 & \cdots & 0 \\
			     0 & 0 & 1 & \cdots & 0 \\
			     \vdots &  & & \ddots  &\vdots \\
			     0 & 0 & \cdots & & -(d-1) \end{array}\right),
\end{equation}
where there are $d-1$ ones on the diagonal so that the matrix is 
traceless and normalized such that Tr$(\lambda_{d^2-1})^2$=2.  
All diagonal matrices have this form and zeros are appended for 
higher dimensions.  We now note that for all diagonal matrices 
Eq.~(\ref{eq:larels}) implies that the result of anti-commutation 
can only produce diagonal matrices.  In other words, the 
$\lambda_k$ in Eq.~(\ref{eq:larels}) are diagonal whenever the left-hand side 
of that equation represents the product of diagonal matrices. Now, noting that 
$$
\{\lambda_i,\lambda_{d^2-1}\} = 2\sqrt{\frac{2}{d(d-1)}} \;\; \lambda_i,
$$
for $i\neq d^2-1$, 
we find that $d_{i,i,d^2-1} = \sqrt{2/d(d-1)}$ for all diagonal 
matrices $\lambda_i$ which are confined to the upper 
$(d-1)\times(d-1)$ block.  There are $d-2$ such matrices since 
there are a total of $d-1$ diagonal matrices.  The only thing left 
to find is the following $d_{d^2-1,d^2-1,i}$.  Let us calculate 
directly using Eq.~(\ref{eq:diagmat}) and Eq.~(\ref{eq:larels}), 
\begin{eqnarray}
\{\lambda_{d^2-1},\lambda_{d^2-1}\} &=& 
                    \frac{4}{d(d-1)}\; \left(\begin{array}{cccc}
                             1 & 0 &  \cdots  & 0 \\
			     0 & 1 &  \cdots & 0 \\
			     \vdots &   & \ddots  &\vdots \\
			     0 & 0 & \cdots  & (d-1)^2 \end{array}\right), \nonumber \\
		    &=& \frac{4}{d}\Bid + 
		            \left(\begin{array}{ccccc}
                             a & 0 & 0 & \cdots  & 0 \\
			     0 & a & 0 & \cdots & 0 \\
			     0 & 0 & a & \cdots & 0 \\
			     \vdots &  & & \ddots  &\vdots \\
			     0 & 0 & \cdots & & b \end{array}\right), \nonumber 
\end{eqnarray}
where $a = 4(2-d)/d(d-1)$ and $b = 4(d-2)/d$.  This implies that 
the only nonzero  $d_{d^2-1,d^2-1,i}$ is 
$$
d_{d^2-1,d^2-1,d^2-1} = -(d-2)\sqrt{\frac{2}{d(d-1)}},
$$
and thus $\sum_id_{iij} = 0$ for SU(d) assuming it is true for 
SU(d-1).  $\square$


\subsection{Exponential of the Sum of Off-diagonals}

\label{app:pod}

Here we show how to calculate the exponential of 
$\sum_{j}\lambda_j \otimes \lambda_j$, we will first show that 
the square of 
$K=\frac{1}{2}\sum_{j\in J}\lambda_j\otimes\lambda_j$ 
is diagonal.  (It is not 
proportional to the identity; but it is diagonal with the same entry 
for each non-zero element on the diagonal.)  

To show this, we write 
the off-diagonal elements in one of the following forms, 
\begin{equation}
\dmat{k}{l}+\dmat{l}{k}, \;\; \mbox{or} \;\; i\dmat{k}{l}-i\dmat{l}{k}.
\end{equation}
Note that $K=\sum_{k<l}Q_{k,l}$
so that $K$ becomes
\begin{eqnarray}
K   &=& 
        \frac{1}{2}\sum_{k<l} [(\dmat{k}{l}+\dmat{l}{k})\otimes(\dmat{k}{l}+\dmat{l}{k})\nonumber \\
        && \;\;\;\;\;\;\;\;\;
            + (i\dmat{k}{l}-i\dmat{l}{k})\otimes (i\dmat{k}{l}-i\dmat{l}{k})] \nonumber \\
    &=&\frac{1}{2} 
         \sum_{k<l} [(\dmat{k}{l}+\dmat{l}{k})\otimes(\dmat{k}{l}+\dmat{l}{k}) \nonumber \\
       && \;\;\;\;\;\;\;\;\; 
            - (\dmat{k}{l}-\dmat{l}{k})\otimes (\dmat{k}{l}-\dmat{l}{k})] \nonumber \\
    &=&\sum_{k<l} [\dmat{k}{l}\otimes\dmat{l}{k}+\dmat{l}{k}\otimes \dmat{k}{l}].
\end{eqnarray}
To square this, we calculate
\begin{eqnarray}
K^2 &=& \sum_{p<q,k<l} [\dmat{p}{q}\otimes\dmat{q}{p}+\dmat{q}{p}\otimes\dmat{p}{q}]\nonumber\\
    &&        \;\;\;\;\;\;\;\;\;        \times
                       [\dmat{k}{l}\otimes\dmat{l}{k}+\dmat{l}{k}\otimes \dmat{k}{l}]\nonumber\\
    &=& \sum_{p<q,k<l} [\delta_{qk}\delta_{pl}\dmat{p}{l}\otimes\dmat{q}{k}
                       +\delta_{ql}\delta_{pk}\dmat{p}{k}\otimes \dmat{q}{l}] \nonumber \\
    &&       \;\;\;\;\;\;\;\;\;         
		       +\delta_{ql}\delta_{pk}\dmat{q}{l}\otimes \dmat{p}{k}]
                       +\delta_{pl}\delta_{qk}\dmat{q}{k}\otimes \dmat{p}{l}]\nonumber
\end{eqnarray}
However, note that the first term requires $p<q=k<l$ and $p=l$. But 
$p=l$ is impossible 
since $p$ is strictly less than $l$.  Therefore this term must vanish.  Similarly 
for the last term.  This gives 
\begin{equation}
K^2 = \sum_{k<l} [\dmat{k}{k}\otimes\dmat{l}{l}+\dmat{l}{l}\otimes \dmat{k}{k})],
\end{equation}
which is clearly diagonal.  $\square$

In fact, the form is clear.  This matrix has 
ones everywhere on the diagonal except at certain places.  
These places are when the indices coincide, i.e., when $k=l$, since we 
are restricted to have $k$ strictly less than $l$.  For example, 
using the Gell-Mann basis for SU(3), a straight-forward calculation using this 
formula produces $K^2=$diag$\{0,1,1,1,0,1,1,1,0\}$.  

It is now readily shown that 
\begin{equation}
K^3 = K, \; \mbox{which implies}\; K^4=K^2, \; \mbox{etc.}  
\end{equation}
This allows us to analytically calculate the exponential of $K$ 
and thus $e_i$ which differs from $K$ by 
the addition of diagonal terms (see Eq.~(\ref{eq:expod2})) 
and tensor products of the identity,  
\begin{equation}
\exp[-it K] = (\Bid-K^2) +K^2\cos(t) - i K \sin(t),
\end{equation}
and thus we obtain SWAP by a slightly different method.


\subsection{Exponential of the Logical X Operator}

\label{app:xbar}

The objective is to calculate $\exp[-i\alpha \bar{X}]$ and 
$\exp[-i\alpha \bar{Z}]$ analytically.  There are several ways 
that we could do this.  Here we choose what we believe is the most
straight-forward. 

Let us recall the definitions  
\begin{equation}
\bar{X} = \sqrt{\frac{1}{12}} \; (e_1-e_2),
\end{equation}
and 
\begin{equation}
\bar{Z} = \frac{1}{6} \; (e_1+e_2-2e_3),
\end{equation}
where 
\begin{eqnarray}
e_1&=&\sum_i \lambda_0\otimes\lambda_i\otimes\lambda_i, \\
e_2&=&\sum_i \lambda_i\otimes\lambda_0\otimes\lambda_i,
\end{eqnarray}
and 
\begin{equation}
e_3=\sum_i \lambda_i\otimes\lambda_i\otimes\lambda_0.
\end{equation}
The dimensionality of $\Bid$ should be clear from context and 
we will use $\Bid$ both for a $d$-state system, $\Bid_d$ and 
also the identity of a composite system 
$\Bid_d\otimes\Bid_d\otimes \cdots \otimes \Bid_d$.

The following identities may be shown, using the identities in 
\ref{app:sunalg}, 
\begin{equation}
e_i^2 = \frac{4}{d^2}(d^2-1)\Bid-\frac{4}{d}e_i.
\end{equation} 
Products of two have a cyclic property:
\begin{eqnarray}
e_1e_2 &=& \frac{2}{d}\sum_i \lambda_i\otimes\lambda_i\otimes\Bid 
          +i\sum_{ijk}f_{ijk}\lambda_j\otimes\lambda_i\otimes\lambda_k \nonumber  \\
      &&  +\sum_{ijk}d_{ijk}\lambda_j\otimes\lambda_i\otimes\lambda_k \nonumber \\
 &=& \frac{2}{d}e_3 -iF +D,  
\end{eqnarray}
and
\begin{eqnarray}
e_1e_3 &=& \frac{2}{d}e_2 +iF +D,  \\
e_2e_3 &=& \frac{2}{d}e_1 -iF +D. 
\end{eqnarray}
Finally, we note that 
\begin{equation}
e_iD = \left(\frac{4}{d^2}\right)\left(\frac{d^2-4}{d}\right)(e_j+e_k) -\frac{12}{d^2}D,
\end{equation}
where $i=1,2,3$ and $i\neq j\neq k\neq i$.

At this point the calculation proceeds in a straight-forward albeit 
tedious manner. One simply computes 
$\bar{X}^3 = [(\sqrt{1/12})(e_1-e_2)]^3$ and 
$\bar{Z}^3 = [(1/6)(e_1 + e_2 - 2e_3)]^3$ 
using the identities in this appendix as well as those in 
\ref{app:sunalg}.  After showing $\bar{Z}^3=\bar{Z}$, 
we know that $\bar{Z}^4=\bar{Z}^2$ (and similarly for $\bar{X}$) 
so the series may be summed to obtain the desired analytic 
expressions for the associated unitary transformations.




\begin{thebibliography}{100}
\bibitem{Kempe:00-fix}
{We note that the corresponding Eq.~(74) in Ref.~\cite{Kempe:00} contains
  incorrect numerical factors, which we have fixed here.}
\bibitem{note:basis}
{Note that the basis being used here is for the set of Gell-Mann matrices. For
  various physical systems, a basis is chosen which corresponds to the
  set of ``good quantum numbers'' for the particular system}.
\bibitem{dgatenote1}
Actually the equation also holds for SU(2) with $d_{ijk}=0$. However, the
  nonzero $d_{ijk}$ make all the difference.
\end{thebibliography}
\end{document}